\documentclass[%
 reprint,
 amsmath,amssymb,
 aps,
]{revtex4-2}

\usepackage{graphicx}
\usepackage{dcolumn}
\usepackage{bm}

\usepackage{newtxmath}
\usepackage{cancel}
\usepackage{tikz}
\newcommand*\circled[1]{\tikz[baseline=(char.base)]{
		\node[shape=circle,draw,inner sep=2pt] (char) {#1};}}
\usepackage{hyperref}
\begin{document}

\preprint{APS/123-QED}

\title{Trapping-loss transition via a saddle-node bifurcation in thermophoretic particle transport driven by a time-periodic vortex}

\author{Srikumar Warrier}
 \email{srikumarw@alum.iisc.ac.in}
\affiliation{%
 Department of Applied Mechanics and Biomedical Engineering \\ Indian Institute of Technology Madras, Chennai 600036, India. 
}%


%

\date{\today}

\begin{abstract}
The transport of inertial particles in unsteady flows is often governed by the competition between multiple migration mechanisms. We investigate the interplay between inertia-induced drift and thermophoretic migration in a time-periodic vortex containing a localized temperature field. Starting from the small-Stokes-number limit of the particle equations of motion, we derive a cycle-averaged radial migration model describing the slow evolution of suspended particles over timescales much longer than the forcing period. The competition between outward inertia-induced drift and inward thermophoretic migration gives rise to stable and unstable fixed points of the reduced radial dynamics, corresponding respectively to particle trapping states and separatrices bounding trapped trajectories. The existence and location of these states are shown to be governed by the dimensionless control parameter $\Pi$, which measures the relative strength of inertia-induced transport to the thermophoretic transport. As $\Pi$ is decreased below a critical value, the stable and unstable fixed points coalesce and disappear, resulting in the loss of particle trapping. Phase portraits, bifurcation diagrams, and local asymptotic analysis demonstrate that trapping is destroyed through a saddle-node bifurcation. The transition is further characterized by the vanishing of the dominant eigenvalue and the associated divergence of the relaxation time, indicative of critical slowing down. Additional calculations employing various other velocity and temperature profiles demonstrate that the trapping-loss mechanism is robust and not specific to a particular profile.

\end{abstract}

\maketitle


\section{Introduction}

The transport of suspended particles in unsteady flows is of fundamental importance in a wide range of natural and engineered systems \citep{balachandar2010turbulent,elghobashi2019direct}, including aerosol transport \cite{michaelides2006particles}, particulate and droplet combustion \cite{williams2013combustion}, and acoustically driven particle manipulation \cite{jethani2026oscillatory}. Unlike passive tracers, inertial particles do not instantaneously adjust to changes in the surrounding fluid motion but instead respond over a finite relaxation time \cite{Maxey1983,Gatignol1983,balachandar2010turbulent,Toschi2009}. Consequently, particle trajectories generally differ from fluid trajectories, breaking the kinematic correspondence between particle and fluid motion. This departure gives rise to transport mechanisms that are absent for passive tracers, including preferential concentration, clustering, migration across streamlines, and long-time drift \cite{Eaton1994,Toschi2009,Bec2007}.

A particularly important class of transport phenomena arises in oscillatory and periodically forced flows. Even when the underlying flow has no net mean motion, time-periodic forcing can produce finite particle transport through nonlinear interactions between particle inertia and oscillatory carrier flow \cite{Lighthill1978,Riley2001,Cartwright2010}. Repeated particle-fluid velocity mismatches may accumulate over many forcing cycles, giving rise to a slow cycle-averaged drift over timescales much longer than the oscillation period \cite{Maxey1983,Toschi2009}. Such drift mechanisms underlie a broad range of phenomena, including acoustic streaming, Stokes drift, and inertia-induced migration in oscillatory vortical flows \cite{Lighthill1978, Riley2001}.

The dynamics of small inertial particles are commonly described within the Maxey--Riley--Gatignol (MRG) framework \cite{Maxey1983,Gatignol1983}, which accounts for the hydrodynamic forces acting on a particle immersed in a non-uniform and time-dependent flow. 
An important consequence of particle inertia is preferential concentration, whereby particles heavier than the fluid are centrifuged out of vortical regions and accumulate in regions of high strain \cite{Eaton1994,Toschi2009}. Such inertial effects strongly influence particle transport and accumulation in dispersed multiphase flows.
Coherent vortical structures are ubiquitous in natural and engineering flows, arising in wake flows, rotating machinery, and swirl-stabilized combustors \citep{Syred1974,Huang2009}. The interaction of inertial particles with such structures has therefore attracted considerable attention. Previous studies have shown that oscillatory and time-dependent vortices can induce long-time particle migration, trapping, and preferential accumulation through cycle-averaged inertial effects \cite{Sapsis2013,Cartwright2010,Rubin2012}. However, these studies have primarily focused on isothermal flows and have largely neglected the influence of thermal gradients. In many practical systems, however, vortical motions coexist naturally with strong thermal gradients. Examples include soot transport in swirl combustors, aerosol dynamics near flames, and acoustically forced thermal flows \citep{manzello2000burning,kadota1984soot}. In such situations, inertia-induced drift and thermophoretic migration act simultaneously, and their competition may fundamentally alter particle transport pathways.

Another important mechanism capable of producing directed particle transport is thermophoresis, whereby suspended particles migrate in response to a temperature gradient \cite{Talbot1980,brock1962,Friedlander2000,Zheng2002,Batchelor1985}. Unlike inertial transport, which arises from the finite response time of particles to an unsteady carrier flow, thermophoresis originates from the asymmetric transfer of momentum from surrounding molecules to the particle surface. When a temperature gradient is present, molecules arriving from the hotter side possess, on average, greater kinetic energy than those arriving from the colder side. The resulting imbalance in molecular impacts generates a net force that drives particles toward regions of lower temperature \citep{Talbot1980,brock1962}. Classical theories of thermophoretic motion predict a drift velocity proportional to the imposed temperature gradient \citep{healy2010experimental}, commonly expressed as

\begin{equation}
\mathbf{V}_T=-K_T \nabla \ln T,
\end{equation}

where $K_T$ is a thermophoretic mobility coefficient that depends on the particle and fluid properties. Thermophoresis plays an important role in aerosol transport, particulate deposition on heated or cooled surfaces, combustion and flame-particle interactions, and atmospheric particulate dynamics. Classical theoretical descriptions were developed by \cite{Talbot1980} and \cite{brock1962}, and further modified by \cite{Beresnev1995,rosner1991size,healy2010experimental}. In combustion systems, thermophoresis has been shown to strongly influence soot transport and deposition in the vicinity of flames \cite{kadota1984soot,manzello2000burning}.

A particularly important feature of thermophoretic transport is that it can generate substantial particle migration even in otherwise quiescent fluids, providing a robust mechanism for directed motion in the absence of a mean flow. Unlike oscillation-induced drift, thermophoretic migration is persistent and does not rely on temporal asymmetries in the carrier flow. Consequently, when thermal gradients coexist with unsteady fluid motion, thermophoretic drift acts alongside inertia-induced transport mechanisms arising from the carrier flow. The resulting particle dynamics are therefore governed by the competition between oscillation-induced excursions and persistent thermophoretic migration, giving rise to rich transport behaviour and non-trivial particle accumulation patterns.

Despite the extensive literature on particle migration in oscillatory flows and thermophoretic transport in thermal gradients, the combined action of these mechanisms remains relatively unexplored. In particular, the manner in which oscillation-induced inertial drift competes with thermophoretic migration to produce particle accumulation, trapping or escape has received comparatively little attention. Understanding this interplay is important because the balance between the two transport mechanisms can fundamentally alter the long-time particle dynamics and give rise to trapping transitions that are absent when either mechanism acts in isolation.

From a dynamical-systems viewpoint, the competition between oscillation-induced drift and thermophoretic migration may generate attracting and repelling states in the effective particle dynamics. Whether such states exist, how they evolve with forcing, and how trapping is ultimately lost are presently unknown. Addressing these questions requires identifying the underlying bifurcation structure governing particle transport.

The present work considers a minimal model consisting of a prescribed oscillatory swirling flow and a localized thermal field. The oscillatory flow generates an outward inertial drift of small particles, while the thermal field induces an opposing thermophoretic migration. Although motivated in part by particle-laden reacting flows, the formulation is intentionally generic and may be viewed as a reduced description of a broader class of periodically forced vortical flows containing localized temperature gradients. The emphasis is therefore placed on the underlying transport mechanisms rather than on a particular engineering configuration.

Within the present framework, we examine how the competition between oscillation-induced inertial drift and thermophoretic migration modifies the long-time transport of suspended particles in periodically forced vortical flows. We show that this competition gives rise to distinct trapping and escape regimes, which admit a natural dynamical-systems interpretation in terms of the effective radial migration dynamics. In particular, trapping is lost through a saddle-node bifurcation as the relative strengths of the competing transport mechanisms are varied. The resulting framework provides a unified description of particle accumulation and escape in thermally stratified
oscillatory flows. The robustness of the proposed mechanism
to variations in the prescribed velocity and temperature
profiles is also examined.

The paper is organized as follows. Section II introduces the governing equations and derives the reduced radial migration model. Section III examines the existence and stability of trapping states. Section IV characterizes the trapping transition and its dependence on the thermophoretic forcing. Section V investigates the associated bifurcation structure and relaxation dynamics. Finally, conclusions are presented in Section VI.

\section{Problem formulation}
\label{sec:problem_formulation}
\begin{figure*}
\includegraphics[width=0.8\linewidth]{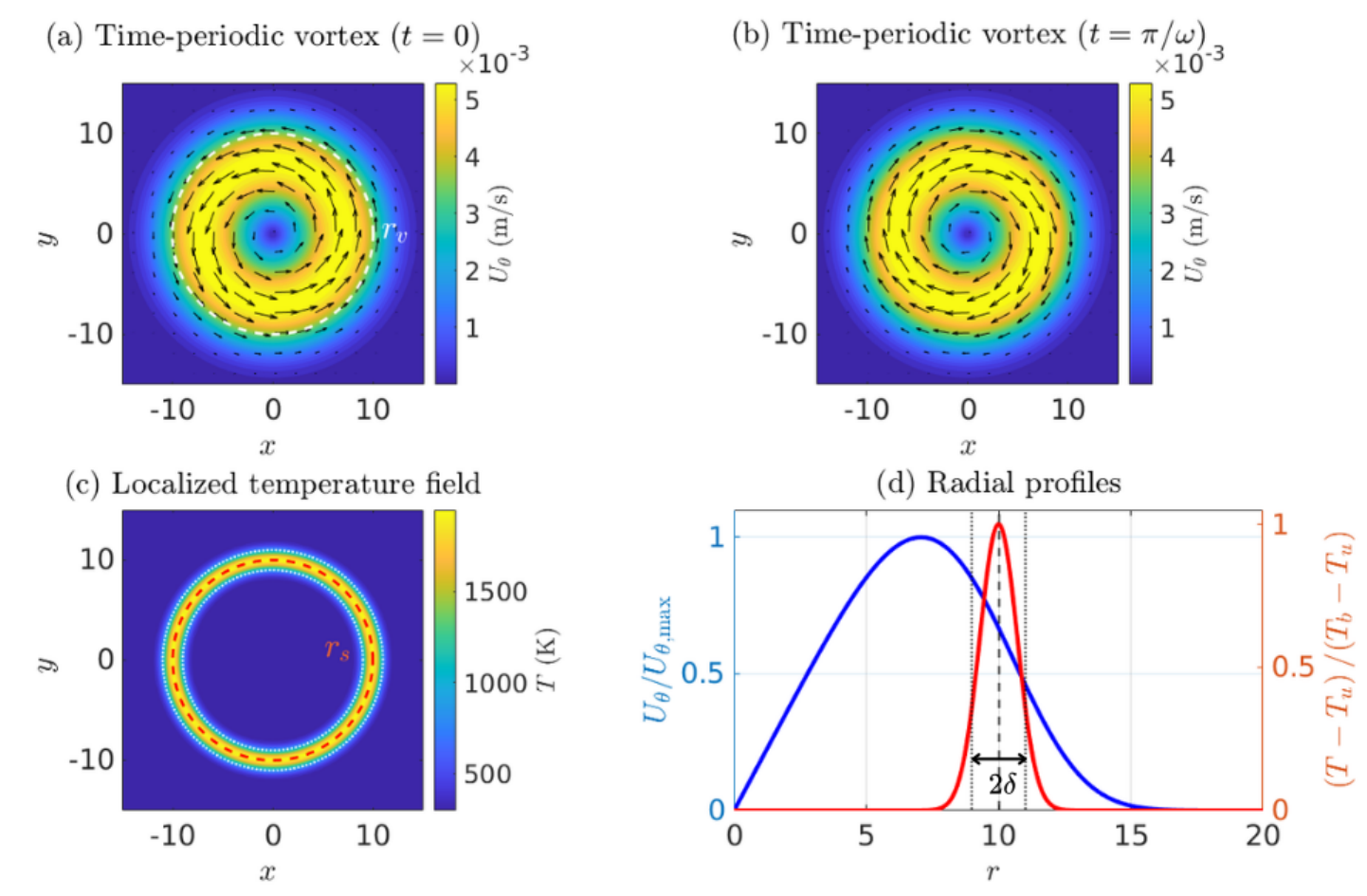}
\caption{\label{fig:wide}(a) shows the time-periodic vortex flow with the velocity vectors (counterclockwise direction) at $t=0$. The parameter $r_v$ denotes
	the characteristic vortex-core radius that sets the radial extent of
	the azimuthal velocity distribution. (b) time-periodic vortex flow with the velocity vectors (counterclockwise direction) at a later time instant $t=\pi/\omega$, illustrating the periodic reversal of the vortex strength. (c) Localized temperature field modeled by an envelope flame with the core region cooler ($T_u$) than the hotter annular region ($T_b$). The parameter $r_s$ denotes the radial location of the temperature
	maximum (flame location) shown as red dashed curve.  (d)  Radial profiles of the azimuthal velocity amplitude $U_\theta(r)$ (in blue) and temperature $T(r)$ (red). The parameter $\delta$ denotes the thermal layer thickness, such that the effective width of the heated annulus is approximately $2\delta$. For the above schematic we choose $T_u=300~\mathrm{K}$, $T_b=2000~\mathrm{K}$, $r_s=10~\mathrm{mm}$, $\delta=1~\mathrm{mm}$, $q=1~\mathrm{s}^{-1}$, $r_v=10~\mathrm{mm}$. }
\label{fig:schematic}
\end{figure*}

The present work employs a reduced-order model designed to isolate the competition between inertial and thermophoretic particle migration.
We consider a prescribed time-periodic vortical flow field together with a localized thermal field, with the thermal structure modeled by an annular temperature distribution representative of an envelope flame as shown in Fig.(\ref{fig:schematic}).
 
Our time-periodic azimuthal velocity component is prescribed as,
\begin{equation}
\mathbf{u}
=
u_{\theta}(r,t)\,\mathbf{e}_{\theta},
\end{equation}

\begin{equation}
u_{\theta}(r,t)
=
U_{\theta}(r)\cos(\omega t),
\end{equation}
$u_{\theta}(r,t)$ is the instantaneous azimuthal velocity component. $r$ is the radial coordinate, $\omega$ is the angular forcing frequency, and $U_{\theta}$
is the amplitude of the velocity given by,

\begin{equation}
U_{\theta}(r)
=
q r
\exp\left[
-\left(
\frac{r}{r_v}
\right)^a
\right],
\label{eq:Utheta}
\end{equation}

where $q$ is the measure of the vortex strength, $r_v$ is the characteristic vortex-core radius, $a$ controls the radial decay of the vortex and therefore determines the shape of the azimuthal velocity profile.

The envelope flame is modeled through the prescribed temperature field $T(r)$,

\begin{equation}
T(r)
=
T_u
+
(T_b-T_u)
\exp
\left[
-\frac{(r-r_s)^2}{\delta^2}
\right],
\label{eq:Tprofile}
\end{equation}

where $T_{u}$ is the ambient gas temperature, $T_{b}$ is the peak temperature in the localized heat source(in the flame), $r_s$ is the radial location of the envelope flame, $\delta$ is the thickness of the thermal layer. The temperature field is assumed to be prescribed and unaffected by particle motion. This approximation is appropriate for dilute suspensions in which particle loading is sufficiently small that thermal feedback from the dispersed phase may be neglected.

Using $r_v$, $q r_v$, $1/q$ and $T_u$ as the characteristic length, velocity, time and temperature scales respectively, we introduce the dimensionless variables,

\begin{equation}
R=\frac{r}{r_v},
\qquad
\tau=qt,
\qquad
\Theta=\frac{T}{T_u},
\qquad \\
R_s=\frac{r_s}{r_v},
\qquad
\Delta=\frac{\delta}{r_v}.
\label{eq:nondimvars}
\end{equation}
 which makes Eq.(\ref{eq:Utheta}),
\begin{equation}
U_{\theta}(R) = R \exp (-(R)^{a})
\label{eq:azimuthal_profile}
\end{equation}

and Eq.(\ref{eq:Tprofile}),
\begin{equation}
\Theta(R)
=
1
+
A_T
\exp\!\left[
-\frac{(R-R_s)^2}{\Delta^2}
\right],
\label{eq:Theta_profile}
\end{equation}

where

\begin{equation}
A_T
=
\frac{T_b-T_u}{T_u}
\label{eq:AT_def}
\end{equation}

is the dimensionless thermal amplitude.

The motion of small inertial particles is described using the reduced Maxey--Riley--Gatignol equation \citep{Maxey1983,Gatignol1983} along with the thermophoretic term in the small-Stokes-number limit ($St\ll1$),

\begin{equation}
\mathbf{v}^{p}
=
\mathbf{u}
-
St
\frac{D\mathbf{u}}{Dt}
+
\mathbf{v}_{th},
\label{eq:MR}
\end{equation}

where Stokes number $St=q\tau_p$ and $\tau_p$ is the particle relaxation time. The thermophoretic drift velocity is modeled as

\begin{equation}
\mathbf{v}_{th}
=
-K_t \nabla \ln \Theta,
\label{eq:vth}
\end{equation}

where $K_t$ denotes the thermophoretic transport coefficient. 
The present formulation considers  micron-sized particles for which deterministic transport mechanisms dominate stochastic Brownian motion. For a particle of radius $a_p\sim 1\upmu\mathrm{m}$ suspended in air at $T=300$K, the Brownian diffusivity given by the Stokes-Einstein relation $D_B=k_B T/6\pi\mu a_p$ where the Boltzmann constant $k_B=1.38\times 10^{-23} \mathrm{JK^{-1}}$, yields  $D_B\sim10^{-11}\mathrm{m^2 s^{-1}}$, whereas typical thermophoretic drift velocities in combustion environments are $O(10^{-3}-10^{-2})\mathrm{ms^{-1}}$ (see \cite{mensch2019measurements}). The corresponding thermophoretic P\'eclet number $Pe_{th}=V_{th}r_v/D_B \sim 10^{6}\gg 1$, for typical values considered here. Brownian diffusion is therefore neglected, and particle transport is assumed to be governed entirely by deterministic inertial and thermophoretic mechanisms.

For a purely azimuthal carrier flow ($U_{\theta}(R)$), the only non zero contribution of the inertial term is the centripetal acceleration term. The radial component of Eq.(\ref{eq:MR}) becomes,
\begin{equation}
 \frac{dR}{d\tau}= St\left[\frac{U_{\theta}^{2}}{R}\cos ^{2}(\Omega\tau)\right] - \left[\Lambda \frac{d(\ln \Theta) }{dR}\right],   \label{eq:radial_MR}
\end{equation}
where,
\begin{equation}
\Omega=\frac{\omega}{q}, 
\qquad
\Lambda=\frac{K_t}{q r_v^{2}},
\end{equation}

$\Omega$ and $\Lambda$ are the dimensionless forcing frequency and the dimensionless thermophoretic transport coefficient respectively.
The present analysis focuses on the long-time migration of particles in a rapidly oscillating flow. We assume that the radial particle displacement occurring over a single forcing cycle is small compared to the characteristic length scale of the vortex. This implies a separation between the fast oscillatory timescale associated with the forcing and the slow timescale associated with radial particle migration.

The forcing timescale is,

\begin{equation}
t_f=\frac{1}{\omega}.
\end{equation}

The characteristic migration timescale associated with the inertia-induced drift is,

\begin{equation}
t_m \sim \frac{1}{St\,q}.
\end{equation}

The ratio of these two timescales defines the asymptotically small parameter,

\begin{equation}
\varepsilon=\frac{t_f}{t_m}
=\frac{St\,q}{\omega}\ll 1.
\label{eq:epsilon}
\end{equation}

Since $\varepsilon\ll1$, particles undergo approximately
$O(\varepsilon^{-1})$ oscillation cycles before
experiencing an $O(1)$ radial displacement, thereby
justifying the introduction of separate fast and slow
timescales. To exploit this separation, we introduce the fast and slow times
\begin{equation}
\tau_s=\varepsilon \tau,
\end{equation}

where $\tau$ describes the rapid oscillatory motion and $\tau_s$ characterizes the slow evolution of the particle position.

The present analysis assumes forcing frequencies
comparable to the characteristic vortex turnover rate,
so that the dimensionless forcing frequency satisfies
\begin{equation}
\Omega=\frac{\omega}{q}=O(1).
\end{equation}
Under this assumption, the timescale separation parameter $\varepsilon=St/\Omega$ satisfies $\varepsilon\sim O(St)$ in the low-Stokes-number limit ($St\ll1$). Consequently, $\varepsilon$ and $St$
become asymptotically equivalent small parameters, and
the multiple-scale expansion may therefore be expressed
directly in powers of $St$. We further note that the high-frequency regime $O(\Omega)\gg1$, where $\epsilon$
and $St$ constitute independent small parameters,
would require a different two-parameter asymptotic
treatment and are beyond the scope of the present work.


The radial particle position is expanded as
\begin{equation}
R(\tau,\tau_s)
=
R_0(\tau_s)
+
St R_1(\tau,\tau_s)
+
O(St^2),
\label{eq:Rexpansion}
\end{equation}

where $R_0$ denotes the slowly varying mean radial position and $R_1$ represents the oscillatory correction induced by the periodic forcing. Since the forcing is periodic, $R_1$ is required to be periodic in the fast time $\tau$,

\begin{equation}
R_1(\tau+2\pi/\Omega,\tau_s)
=
R_1(\tau,\tau_s),
\label{eq:periodic}
\end{equation}

with zero mean over one forcing cycle,

\begin{equation}
\left\langle R_1 \right\rangle =0.
\end{equation}

Substituting Eq.(\ref{eq:Rexpansion}) into Eq.(\ref{eq:radial_MR}), expanding the drift terms about the slowly varying mean position $R_0$, and collecting terms order-by-order in $\varepsilon$ yields an evolution equation for the mean particle position. The details of this multiple-timescale analysis are presented in  Appendix \ref{appdx:Multiscale formulation}.

Averaging over one forcing period eliminates all oscillatory contributions and yields the evolution equation for the slow radial position,

\begin{equation}
\frac{dR_0}{d\tau_s}
=
St
\left\langle
\frac{U_\theta^2(R_0,\tau)}{R_0} \cos^{2}(\Omega\tau)
\right\rangle
-
\Lambda
\frac{d\ln\Theta}{dR_0}.
\label{eq:R0slow}
\end{equation}

For the prescribed oscillatory vortex

\begin{equation}
u_\theta(R,\tau)
=
U_\theta(R)\cos(\Omega\tau),
\end{equation}

the cycle average satisfies

\begin{equation}
\left\langle
\cos^2(\Omega\tau)
\right\rangle
=
\frac{1}{2},
\end{equation}

which yields

\begin{equation}
\left\langle
\frac{D\mathbf{u}}{Dt}
\right\rangle
=
-\frac{U_\theta^2(R_0)}
{2R_0}\,
\mathbf{e}_r.
\label{eq:avgacc}
\end{equation}

The slow migration equation therefore becomes


\begin{equation}
    \frac{dR_{0}}{d\tau_{s}}=\frac{U_{\theta}^{2}\left(R_{0}\right)}{2R_{0}}-\Pi\left(\frac{d\ln\Theta}{dR_0}\right),
    \label{eq:slowdrift}
\end{equation}
where,
\begin{equation}
\Pi
=
\frac{\Lambda}{St}
=
\frac{K_t}
{\tau_p q^2 r_v^2}.
\label{eq:Pi}
\end{equation}
Eq.(\ref{eq:slowdrift}) is the slow evolution of the cycle-averaged particle position in the slow timescale. Here $R_0$ denotes the cycle-averaged particle position evolving on the slow timescale $\tau_s$. For convenience, the slow evolution equation is re-expressed in terms of $\tau$ and we drop the subscript $0$. The non-dimensional drift velocity equation is,
\begin{equation}
\frac{dR}{d\tau}=St\frac{U_{\theta}^{2}(R)}{2R}-\Lambda\frac{d\ln \Theta}{dR}.
\label{eq:driftdim}
\end{equation}

Eq.(\ref{eq:driftdim}) describes the competition between radially outward inertial migration induced by the time-periodic vortex and thermophoretic migration driven by the temperature field. 
Substituting the prescribed velocity field (Eq.~(\ref{eq:azimuthal_profile})) and temperature field (Eq.(\ref{eq:Theta_profile})) into Eq.~(\ref{eq:driftdim}) along with Eq.(\ref{eq:nondimvars}) yields, the radial drift equation for the present velocity and temperature field,

\begin{equation}
\frac{dR}{d\tau}
=
\frac{St}{2}
R e^{-2R^a}
-
\Lambda
\frac{d\ln\Theta}{dR}.
\label{eq:driftnondim}
\end{equation}

 Particle trapping corresponds to equilibrium solutions of Eq.~(\ref{eq:driftnondim}), obtained from,

\begin{equation}
\frac{dR}{d\tau}=0.
\label{eq:trap}
\end{equation}

Dividing Eq.~(\ref{eq:driftnondim}) by $St$ shows that the equilibrium structure depends only on the ratio $\Pi=\Lambda/St$. The control parameter $\Pi$ measures the relative importance of thermophoretic and inertial transport. Large values of $\Pi$ correspond to thermophoresis-dominated dynamics, whereas small values correspond to inertia-dominated transport. Experimentally, $\Pi$ may be varied through changes in the forcing strength $q$, particle relaxation time $\tau_p$, or thermophoretic mobility $K_t$.

The fixed-point condition may therefore be written as

\begin{equation}
F(R;\Pi)
=
\frac{1}{2}
R e^{-2R^a}
-
\Pi
\frac{d\ln\Theta}{dR}
=
0.
\label{eq:fixed_point_equation}
\end{equation}
Eq.(\ref{eq:fixed_point_equation}) recasts the particle-transport problem as a one-dimensional nonlinear dynamical system parameterized by the control parameter $\Pi$. Within this framework, particle trapping corresponds to stable fixed points of the reduced radial dynamics, while particle escape occurs when no such attracting states exist. The central question is therefore, \textit{``how does the fixed-point structure of $F(R;\Pi)$ evolve as the relative strengths of thermophoretic and inertia-induced transport are varied through $\Pi$?"} In particular, we seek to determine the conditions under which stable trapping states exist, whether they are accompanied by unstable equilibria that bound the trapping region, and how these states are ultimately destroyed as the balance between the competing transport mechanisms is altered. 
\newline
In the following section (section \ref{sec:results_and_discussion}), we address these questions by examining the existence and stability of the fixed points of Eq.(\ref{eq:fixed_point_equation}), characterizing the trapping-loss transition, and identifying the dynamical mechanism responsible for particle escape.

\section{Results and discussion}
\label{sec:results_and_discussion}
The time-averaged particle dynamics is governed by the competition between the outward inertial drift generated by the oscillatory vortex and the thermophoretic drift induced by the localized thermal field. We begin by examining the structure of these individual drift mechanisms and demonstrate how their balance gives rise to particle trapping states. We then characterize the stability of these states, identify the trapping-loss transition, and show that particle escape occurs through a saddle-node bifurcation accompanied by critical slowing down.

\subsection{Competing transport mechanisms and trapping-loss transitions}
The dynamics of particle migration is governed by the drift function $F(R;\Pi)$ appearing in Eq.~(\ref{eq:fixed_point_equation}). The two terms comprising $F$ represent distinct transport mechanisms: the first term corresponds to the cycle-averaged inertia-induced drift generated by the oscillatory vortex, while the second term represents thermophoretic migration driven by the localized temperature gradient. Particle trapping is therefore determined by the balance between these competing contributions.

\begin{figure*}[ht!]
\includegraphics[width=1.07\linewidth]{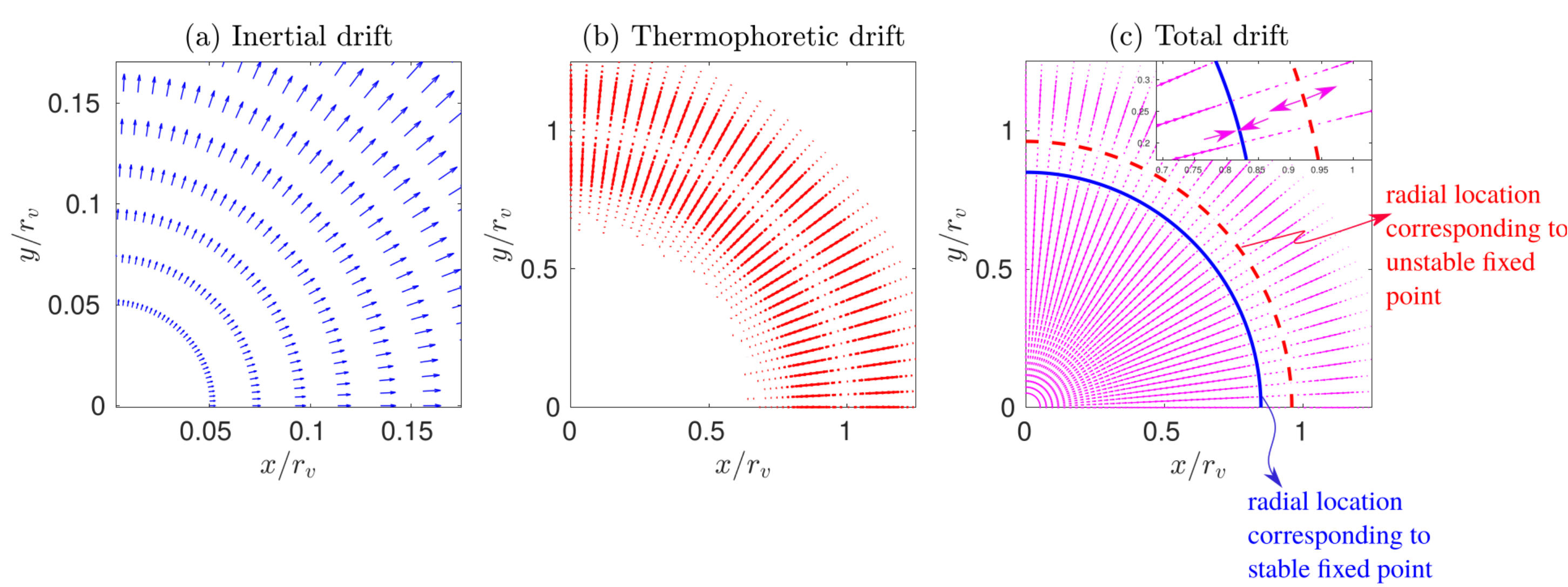}
\caption{\label{fig:wide} Time-averaged particle drift mechanisms. (a) Inertial drift generated by the oscillatory vortex, directed radially outward. (b) Thermophoretic drift induced by the localized thermal field, directed inward toward colder regions. (c) Total drift field obtained from the superposition of the two mechanisms. Inset: Enlargement of the vicinity of the trapping region. The radial drift velocity changes sign at two locations corresponding to the stable (solid blue) and unstable (dashed red) fixed points of the reduced radial dynamics. Perturbations about the stable fixed point are restored toward equilibrium, leading to particle trapping, whereas the unstable fixed point is repelling and forms the boundary of the trapping basin. The coexistence of these fixed points arises from the competition between outward inertial drift and inward thermophoretic migration}
\label{fig:competing_drift}
\end{figure*}

Figures~\ref{fig:competing_drift}(a,b) show the individual drift components entering Eq.(\ref{eq:fixed_point_equation}). The inertial contribution is everywhere positive and promotes outward migration, whereas the thermophoretic contribution is directed inward within the vicinity of the thermal layer. The existence of trapping states therefore depends on whether these two contributions can balance at finite radius.

\begin{figure}
\centering
\includegraphics[width=0.8\linewidth]{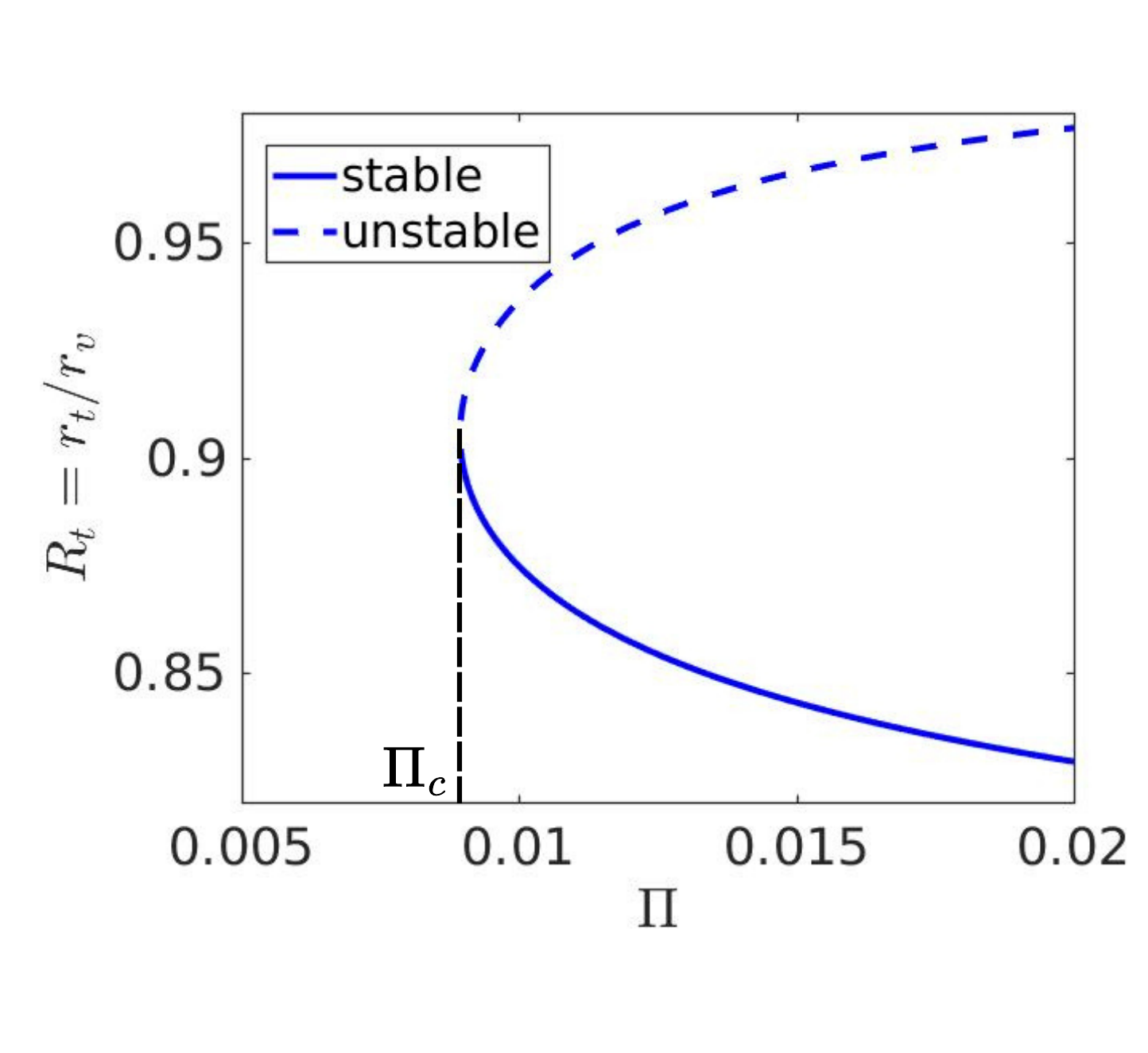}
\caption{Bifurcation diagram of the reduced radial dynamics showing the stable (solid) and unstable (dashed) fixed points as a function of the dimensionless control parameter $\Pi$. The lower branch corresponds to the stable trapping radius, while the upper branch represents the unstable fixed point that bounds the trapping basin. The two branches merge at the critical value $\Pi_c$, beyond which no trapping states exist. The finite-parameter coalescence of the stable and unstable branches is the characteristic signature of a saddle-node bifurcation. For the above figure $a=4$, $R_s=1$, $\Delta=0.1$, $A_T=5.667$ and computed $\Pi_c \approx 0.0089$.}
\label{fig:bifurcation_diagram}
\end{figure}

This balance is illustrated in Fig.~\ref{fig:competing_drift}(c), which shows the total drift field corresponding to $F(R;\Pi)$. Fixed points of the reduced dynamics satisfy $F(R;\Pi)=0$ and occur at locations where the net drift changes sign. The figure reveals the existence of two such equilibria. At the inner fixed point, particles located slightly inside drift outward while particles located slightly outside drift inward, causing trajectories to converge toward the equilibrium location. This fixed point therefore acts as a stable trapping state and is characterized by a negative local slope of the drift function,

\begin{equation}
\left.\frac{\partial F}{\partial R}\right|_{\mbox{stable fp}}<0.
\end{equation}

In contrast, trajectories on either side of the outer equilibrium move away from the fixed point, implying

\begin{equation}
\left.\frac{\partial F}{\partial R}\right|_{\mbox{unstable fp}}>0,
\end{equation}

The unstable fixed point acts as a separatrix between trapped and escaping trajectories and therefore defines the outer boundary of the trapping basin.

The existence of both stable and unstable equilibria demonstrates that trapping is fundamentally a fixed-point phenomenon of the reduced dynamical system. This immediately raises the question of how the solutions of $F(R;\Pi)=0$ (Eq.~(\ref{eq:fixed_point_equation})) evolve as the control parameter $\Pi$ is varied. As shown next, decreasing the value of $\Pi$  causes the stable and unstable branches to approach one another and ultimately disappear through a saddle-node bifurcation. To answer this systematically, we construct a bifurcation diagram by tracking the equilibrium solutions of Eq.~(\ref{eq:fixed_point_equation}) as a function of $\Pi$. 
Figure~\ref{fig:bifurcation_diagram} shows the resulting equilibrium branches. For $\Pi >\Pi_c$, Eq.~(\ref{eq:fixed_point_equation}) admits two distinct solutions corresponding to the stable trapping state and the unstable separatrix identified in Fig.~\ref{fig:competing_drift}. As $\Pi$ is reduced, the two branches approach one another, indicating that the region over which inward thermophoretic migration can balance outward inertial transport progressively shrinks. At a critical value $\Pi=\Pi_c$, the stable and unstable branches coalesce at a single equilibrium location. Mathematically, this point is characterized by the simultaneous satisfaction of the fixed-point condition $F(R;\Pi_c)=0$ and $\frac{\partial F}{\partial R}(R;\Pi_c)=0$.
The critical value $\Pi_c$ is determined numerically in fig.(\ref{fig:bifurcation_diagram}). For each value of $\Pi$, the roots of the drift function $F(R;\Pi)=0$ is located numerically and classified according to the sign of $\frac{dF}{dR}$. The critical parameter $\Pi_c$ is identified as the value of $\Pi$ at which the stable and unstable branches coalesce.
For $\Pi<\Pi_c$, Eq.~(\ref{eq:fixed_point_equation}) possesses no real solutions and the reduced dynamics no longer contains any equilibrium trapping states. The bifurcation diagram therefore identifies $\Pi_c$ as the threshold separating two qualitatively distinct dynamical regimes.  

\begin{figure*}[ht!]
\centering
\includegraphics[width=1.07\linewidth]{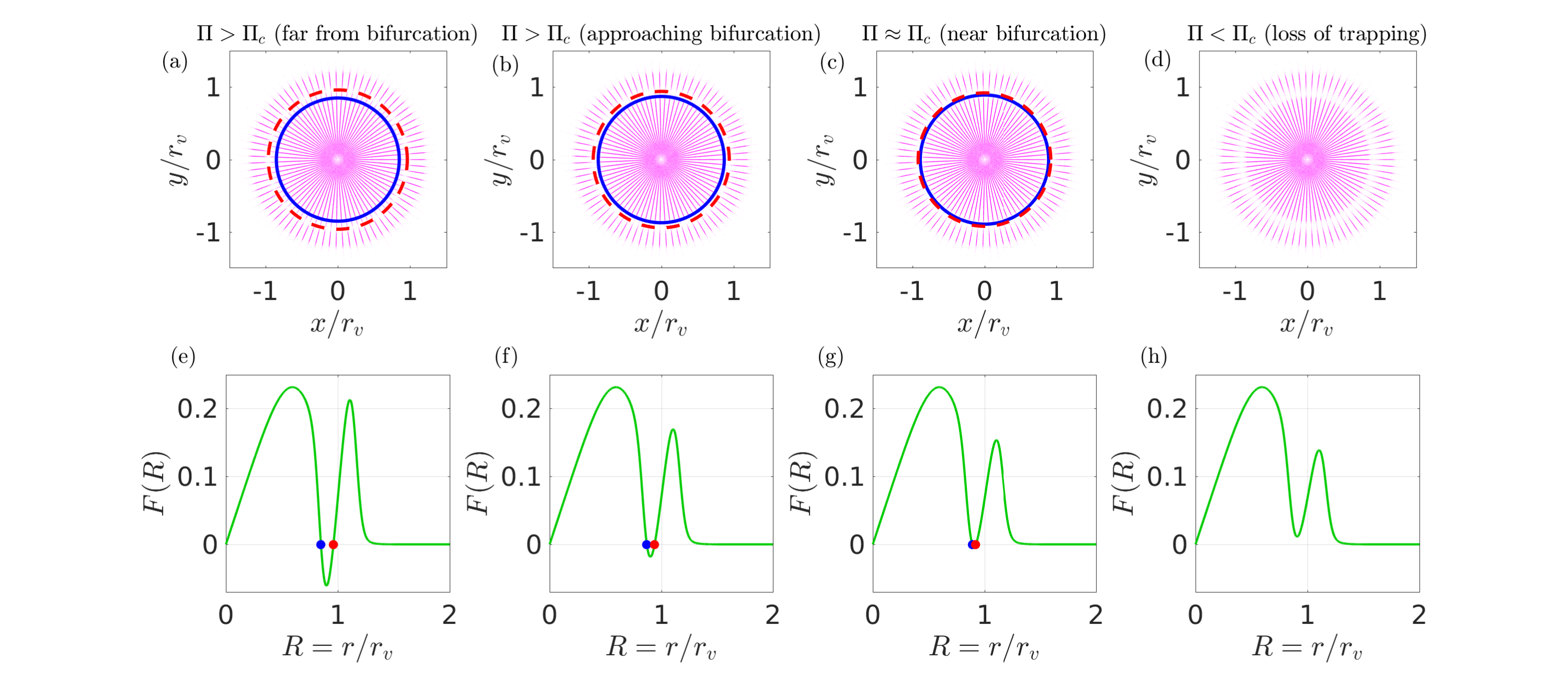}
\caption{Evolution of the reduced radial dynamics as the control parameter $\Pi$ decreases toward the critical value $\Pi_c$. Top row: (a-d) total radial drift field. The solid blue circle denotes the stable trapping radius, while the dashed red circle denotes the unstable fixed point. Bottom row: corresponding drift function $F(R)$, with stable and unstable fixed points indicated by blue and red markers, respectively.  As $\Pi$ approaches $\Pi_c$ (e-g), the stable and unstable fixed points move toward one another and eventually coalesce. For $\Pi<\Pi_c$ (h), no fixed points exist and particle trapping is lost. The drift velocity remains positive throughout the domain. Although the drift becomes
small near the former trapping location, particles are
not trapped and eventually escape. The disappearance of the stable–unstable pair identifies the trapping-loss transition as a saddle-node bifurcation.}
\label{fig:fixed_point_coalescence}
\end{figure*}

Figure~\ref{fig:fixed_point_coalescence} provides a physical-space visualization of this transition by showing the evolution of the drift field and the corresponding radial drift function $F(R)$ in the vicinity of the critical point. The blue and red markers denote the stable and unstable fixed points, respectively, corresponding to the equilibrium solutions of Eq.~(\ref{eq:fixed_point_equation}).

For $\Pi>\Pi_c$, the drift function exhibits two distinct zero crossings associated with the stable trapping state (blue point in the bottom row) and the unstable separatrix (red point in the bottom row). In the upper panels, these equilibria appear as concentric circular loci that bound the trapping region. As $\Pi$ approaches $\Pi_c$, the separation between the blue and red fixed points progressively decreases, indicating a shrinking trapping basin and a weakening of the trapping state. Near the critical point, the two equilibria become nearly indistinguishable. The corresponding drift field shows the stable trapping state and unstable separatrix approaching one another until they eventually collide. Beyond the critical point, the two fixed points disappear simultaneously, and the drift function no longer possesses any zero crossings.  The drift function remains positive
throughout the domain and therefore no equilibrium
trapping states exist. Although the radial drift velocity
becomes very small in the vicinity of the former
bifurcation point, it never changes sign. Particles
therefore continue to migrate outward and ultimately
escape the trapping region.
Consequently, the trapping basin collapses and particles are no longer confined to a finite radial location. 

In the present study, the bifurcation diagram (Fig.(\ref{fig:bifurcation_diagram})), the non-dimensional parameter values considered are $a=4$, $R_s=1$, $\Delta=0.1$, $A_T=5.667$. The control parameter $\Pi$ is varied directly over the range $0.008\leq \Pi \leq 0.02$, which spans the critical value $\Pi_c\approx0.0089$. Since $\Pi=K_t/(\tau_p q^2 r_v^2)$, varying $\Pi$ is equivalent to varying either the thermophoretic transport coefficient $K_t$, particle response time $\tau_p$, vortex strength $q$, or vortex-core radius $r_v$ while keeping the remaining parameters fixed. 

Fig.(\ref{fig:fixed_point_coalescence}) provides a direct geometric interpretation of the bifurcation diagram : the stable trapping state is destroyed through its collision with the unstable separatrix, eliminating the basin of attraction and leaving escape as the only accessible long-time behavior.

\subsection{Local Bifurcation Analysis}
The coalescence of the stable and unstable equilibria observed in Fig.(\ref{fig:bifurcation_diagram}) strongly suggests that trapping is lost through a saddle-node bifurcation. To establish this more formally, we perform a local bifurcation analysis in the vicinity of the critical point $(R_c^{*},\Pi_c)$.
The cycle-averaged particle transport is governed by the one-dimensional dynamical system introduced in Eq.(\ref{eq:fixed_point_equation}) which may be written in the generic form,

\begin{equation}
\frac{dR}{d\tau}=F(R;\Pi),
\label{eq:local_dynamics}
\end{equation}
At the bifurcation point, the stable and unstable fixed points coalesce at a critical equilibrium location $R_c^*$ corresponding to the critical parameter value $\Pi_c$. The fixed-point condition is

\begin{equation}
F(R^*;\Pi)=0,
\label{eq:fp_condition}
\end{equation}

where $R^*$ denotes an equilibrium solution of the reduced radial dynamics. At the bifurcation point, the stable and unstable branches merge and the drift function becomes tangent to the $R$-axis. Consequently, the critical point satisfies

\begin{equation}
F(R_c^*;\Pi_c)=0,
\label{eq:SN1}
\end{equation}

and

\begin{equation}
\left.
\frac{\partial F}{\partial R}
\right|_{(R_c^*,\Pi_c)}
=0.
\label{eq:SN2}
\end{equation}

Equations~(\ref{eq:SN1}) and (\ref{eq:SN2}) are the defining conditions of a saddle-node bifurcation.

To analyze the behavior near the bifurcation, we introduce the local variables

\begin{equation}
x = R-R_c^*,
\label{eq:xdef}
\end{equation}

and

\begin{equation}
\mu = \Pi-\Pi_c,
\label{eq:mudef}
\end{equation}

where $x$ measures the displacement from the critical equilibrium location and $\mu$ measures the distance from the critical parameter value.

The drift function may then be expanded in a Taylor series about the point $(R_c^*,\Pi_c)$,


    \begin{align}
        F(R_c^*+x,\Pi_c+\mu)=F_c +xF_R  + \mu F_{\Pi}
	\\ + \frac{x^2}{2}F_{RR}
	+x\mu F_{R\Pi}
	+ \frac{\mu^2}{2}F_{\Pi\Pi}
	+\cdots ,
       \label{eq:taylor}
    \end{align}
 

where
\begin{equation}
\begin{aligned}
F_R &=
\left(\frac{\partial F}{\partial R}\right)_{(R_c^*,\Pi_c)},
\qquad
F_{\Pi} =
\left(\frac{\partial F}{\partial \Pi}\right)_{(R_c^*,\Pi_c)},\\[2mm]
F_{RR} &=
\left(\frac{\partial^2 F}{\partial R^2}\right)_{(R_c^*,\Pi_c)},
\qquad
F_{\Pi\Pi} =
\left(\frac{\partial^2 F}{\partial \Pi^2}\right)_{(R_c^*,\Pi_c)},\\[2mm]
F_{R\Pi} &=
\left(\frac{\partial^2 F}{\partial R\,\partial \Pi}\right)_{(R_c^*,\Pi_c)}.
\end{aligned}
\end{equation}

Using Eqs.(\ref{eq:SN1}) and (\ref{eq:SN2}), the first two terms vanish, yielding

\begin{equation}
F(R;\Pi)
=
\mu F_{\Pi}
+\frac{x^2}{2}F_{RR}
+x\mu F_{R\Pi}
+\frac{\mu^2}{2}F_{\Pi\Pi}
+\cdots .
\label{eq:taylor_reduced}
\end{equation}

The fixed points of the system satisfy Eq.~(\ref{eq:fp_condition}), and therefore correspond to the roots of Eq.~(\ref{eq:taylor_reduced}). Near the bifurcation, the dominant balance is obtained by equating the first surviving parameter-dependent term and the first surviving spatial term,

\begin{equation}
F_{\Pi}\mu
\sim
\frac{x^2}{2}F_{RR}.
\label{eq:dominant_balance}
\end{equation}

This implies

\begin{equation}
x \sim \mu^{1/2},
\label{eq:sqrt_scaling}
\end{equation}

which is the characteristic square-root scaling associated with saddle-node bifurcations. Since $x\mu=O(\mu^{3/2})$ and $\mu^2=O(\mu^2)$, the mixed and higher-order terms are asymptotically smaller and may be neglected to leading order. Eq.(\ref{eq:taylor_reduced}) therefore reduces to

\begin{equation}
F(R;\Pi)
\simeq
F_{\Pi}\mu
+
\frac{1}{2}F_{RR}x^2.
\label{eq:normalform1}
\end{equation}

Substituting Eq.~(\ref{eq:normalform1}) into Eq.~(\ref{eq:local_dynamics}) and using Eq.(\ref{eq:xdef}),

\begin{equation}
\frac{dR}{d\tau}=\frac{dx}{d\tau}
=
A\mu
+
Bx^2,
\label{eq:normalform}
\end{equation}

where

\begin{equation}
A=F_{\Pi},
\qquad
B=\frac{1}{2}F_{RR}.
\end{equation}

Eq.(\ref{eq:normalform}) is the canonical normal form of a saddle-node bifurcation \cite{strogatz2024nonlinear}. Further the structure of the fixed point can be found by noting that the equilibrium solutions of Eq.(\ref{eq:normalform}) satisfy,

\begin{equation}
A\mu+Bx^2=0.
\label{eq:fp_normalform}
\end{equation}

Solving for $x$ gives

\begin{equation}
x_{\pm}
=
\pm
\sqrt{-\frac{A}{B}\mu}.
\label{eq:xpm}
\end{equation}

Transforming back to the original coordinate yields

\begin{equation}
R_{\pm}^*
=
R_c^*
\pm
\sqrt{-\frac{A}{B}\mu}.
\label{eq:Rpm}
\end{equation}

The separation between the stable and unstable branches is therefore

\begin{equation}
\Delta R
=
R_+^*-R_-^*
=
2
\sqrt{-\frac{A}{B}\mu},
\label{eq:branchsep}
\end{equation}

which implies

\begin{equation}
\Delta R
\propto
(\Pi-\Pi_c)^{1/2}.
\label{eq:branch_scaling}
\end{equation}

Eq.(\ref{eq:branch_scaling}) predicts the characteristic square-root opening of the bifurcation diagram shown in Fig.~\ref{fig:bifurcation_diagram}.
An important dynamical consequence of such bifurcations is the phenomenon of critical slowing down, whereby the recovery of the system from small perturbations becomes progressively slower as the critical point is approached.

The origin of this behavior can be understood by examining the local stability of the trapping state. Since critical slowing down is associated with the long-time evolution of small perturbations about an equilibrium, it is sufficient to linearize the reduced radial dynamics in the vicinity of a fixed point. The resulting eigenvalue governs the asymptotic rate at which perturbations either decay or grow. As the stable and unstable equilibria approach one another near $\Pi_c$, the restoring dynamics weakens, causing the magnitude of the eigenvalue to decrease. At the bifurcation point, the eigenvalue vanishes, implying an infinite recovery time. Consequently, the scaling of the eigenvalue provides direct information about the characteristic relaxation time and the onset of critical slowing down.

\begin{figure}
\centering
\includegraphics[width=1\linewidth]{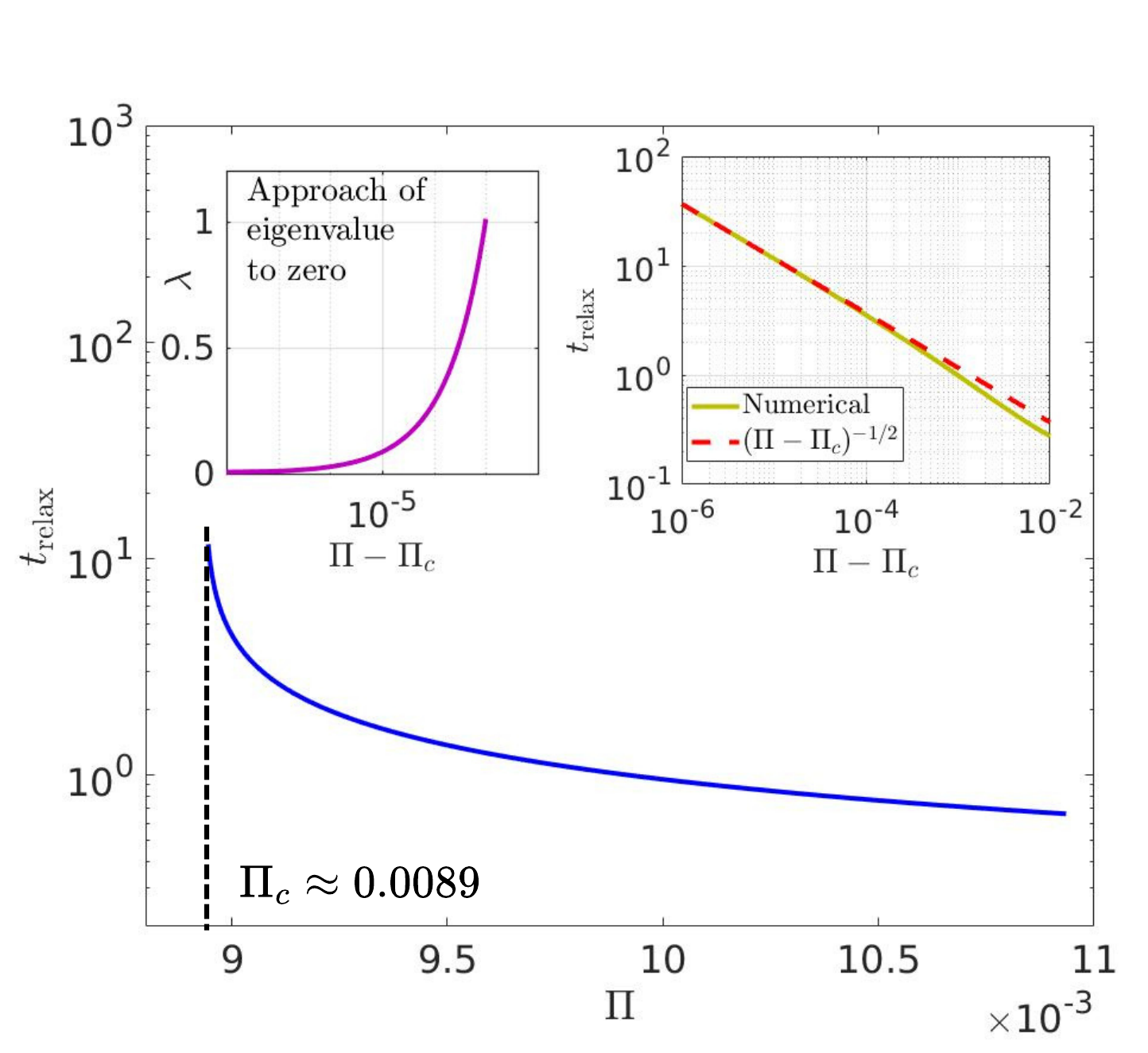}
\caption{Critical slowing down near the trapping-loss transition. The relaxation time $t_{\mathrm{relax}}=1/|\lambda|$ associated with perturbations about the stable trapping state is plotted as a function of the control parameter $\Pi$. As $\Pi$ approaches the critical value $\Pi_c \approx 0.0089$ from above, the relaxation time increases sharply, indicating critical slowing down and the loss of stability at the trapping threshold. The vertical dashed line marks the saddle-node bifurcation point. Inset: (left): shows the corresponding dominant eigenvalue ($\lambda=\left.\partial F/\partial R\right|_{R=R^*}$), which approaches zero as ($\Pi \rightarrow \Pi_c$).   (right) Log-log comparison between the numerical relaxation time and the asymptotic saddle-node prediction $t_{\mathrm{relax}}\sim(\Pi-\Pi_c)^{-1/2}$. The excellent agreement close to $\Pi_c$ confirms the characteristic square-root scaling associated with a saddle-node bifurcation, while deviations farther from the critical point reflect departure from the asymptotic regime. Non-dimensional values for the above figure used $a=4$, $R_s=1$, $\Delta=0.1$, $A_T=5.667$.}
\label{fig:slow_down_approach}
\end{figure}

Figure~\ref{fig:slow_down_approach} examines these predictions by measuring the dominant eigenvalue and the associated relaxation time as the trapping-loss transition is approached. 
To determine the stability of the equilibria, we introduce a small perturbation $\eta$ about a fixed point,

\begin{equation}
R=R^*+\eta,
\qquad
|\eta|\ll1.
\end{equation}

Substituting into Eq.~(\ref{eq:local_dynamics}) and linearizing yields

\begin{equation}
\frac{d\eta}{d\tau}
=
\lambda \eta,
\label{eq:linearized}
\end{equation}

where

\begin{equation}
\lambda
=
\left.
\frac{\partial F}{\partial R}
\right|_{R=R^*}
\label{eq:eigenvalue}
\end{equation}

is the local eigenvalue governing the asymptotic evolution of perturbations.

Using the normal form Eq.(\ref{eq:normalform}),

\begin{equation}
\lambda
=
2Bx.
\label{eq:lambda_x}
\end{equation}

Evaluating Eq.(\ref{eq:lambda_x}) at equilibrium solutions, Eq.(\ref{eq:xpm}) gives,

\begin{equation}
\lambda_{\pm}
=
\pm
2B
\sqrt{-\frac{A}{B}\mu}.
\label{eq:lambdapm}
\end{equation}

Hence,

\begin{equation}
|\lambda|
\propto
(\Pi-\Pi_c)^{1/2}.
\label{eq:eigen_scaling}
\end{equation}

As the bifurcation point is approached, the magnitude of the eigenvalue decreases and vanishes at $\Pi=\Pi_c$. The characteristic relaxation time associated with perturbation decay is,

\begin{equation}
t_{\mathrm{relax}}
=
\frac{1}{|\lambda|},
\label{eq:trelax}
\end{equation}

which immediately yields,

\begin{equation}
t_{\mathrm{relax}}
\propto
(\Pi-\Pi_c)^{-1/2}.
\label{eq:trelax_scaling}
\end{equation}

Eq.(\ref{eq:trelax_scaling}) predicts a divergence of the relaxation time as the bifurcation point is approached, which is the defining signature of critical slowing down.

The results presented thus far have been obtained using the Gaussian profile (Eq.\ref{eq:azimuthal_profile}) and the Gaussian temperature distribution
(Eq.\ref{eq:Theta_profile}). To ascertain that the observed
trapping-loss transition is not an artifact of these particular
choices, Appendix~\ref{appdx:robustness} examines several
alternative velocity and temperature profiles. The calculations
demonstrate that the existence of stable trapping states, their
subsequent annihilation through a saddle-node bifurcation, and
the associated trapping-loss transition remain qualitatively
unaltered. The bifurcation mechanism identified in the present
study is therefore robust to variations in both the vortical
and temperature fields.

\section{Conclusions}
\label{sec:conclusions}

The present study examined the long-time transport of inertial particles subjected simultaneously to thermophoretic migration and oscillation-induced inertial drift in a time-periodic vortex. Although the underlying flow possesses no mean radial motion, cycle-averaged inertial effects generate a finite outward particle drift that competes directly with inward thermophoretic transport. By exploiting the separation between the fast oscillatory timescale and the slow migration timescale, the particle dynamics are reduced to an effective one-dimensional radial drift equation whose fixed points govern particle trapping.

In the present study, we restrict our analysis to particles with small Stokes numbers in order to facilitate analytical progress. For all calculations, we consider a representative $1~\upmu$m soot particle suspended in air. Typical soot particle density $\rho_p \approx 1800~\mathrm{kg\,m^{-3}}$, the corresponding particle relaxation time is estimated as $\tau_p = 2\rho_p a_p^2/9\mu \approx 2\times10^{-5}~\mathrm{s}$. For characteristic vortex strengths in the range $q\sim1$--$10^3~\mathrm{s^{-1}}$, the resulting Stokes number lies in the range $10^{-5}\lesssim St \lesssim 10^{-2}$, thereby satisfying the asymptotic requirement $St\ll1$ underlying the present analysis.

Analysis of the reduced dynamics reveals that particle trapping arises from a balance between outward inertia-induced drift and inward thermophoretic migration. Depending on the relative strengths of the two transport mechanisms, particles either accumulate at a finite radius or escape from the trapping region. The loss of trapping occurs through a saddle-node bifurcation in which stable and unstable fixed points merge and annihilate as the control parameter is varied.

This dynamical interpretation is supported by the fact that bifurcation diagrams exhibit the characteristic coalescence of stable and unstable branches, while the associated eigenvalue approaches zero at the critical point, giving rise to critical slowing down and a divergence of the relaxation time. Local asymptotic analysis in Appendix \ref{appdx:Multiscale formulation} further recovers the canonical saddle-node normal form, thereby providing a theoretical basis for the observed trapping-loss transition.

A central result of the present work is the identification of the dimensionless parameter $\Pi=\Lambda/St$, which quantifies the competition between thermophoretic and inertia-induced transport. The existence and location of trapping states, as well as the onset of trapping loss, are governed primarily by this single parameter. The reduced framework developed here is intentionally generic and may therefore provide insight into a broader class of oscillatory vortical flows containing localized thermal gradients, for example, the particle transport inside annular swirl stabilized combustors, where coherent vortical structures interact with an envelope flame front, and thermally stratified vortex flows generated by localized heating in laboratory or environmental settings.

Several extensions of the present theory remain open. The present study considers micron-sized particles ($a_p\sim1~\upmu$m), for which Brownian diffusion is negligible compared to thermophoretic and inertia-induced transport. However, soot particles encountered in combustion environments are often in the nanometer size range \citep{manzello2000burning,mensch2019measurements}, in which case Brownian motion may become important and significantly alter the trapping dynamics. Furthermore, the present analysis is restricted to dilute suspensions in which particle--particle interactions are neglected. The inclusion of hydrodynamic interactions in moderately concentrated systems may lead to non-trivial modifications to the trapping states. Finally, the present framework considers the reduced radial dynamics, with the particle evolution depending only on its instantaneous state. In unsteady flows such as those considered here, the Basset history force may become important by introducing memory effects associated with the particle's past motion \cite{prasath2019accurate}. Such non-Markovian dynamics could qualitatively modify the trapping states and the saddle-node trapping-loss transition identified in the present study.

\begin{acknowledgments}
The author wishes to acknowledge  Dr. Anubhab Roy at the Department of Applied Mechanics, IIT Madras, and the Institute Postdoctoral Fellowship (IPDF) program at IIT Madras for the support and encouragement. 
\end{acknowledgments}

\appendix
\section{Appendix: Multiscale formulation}
\label{appdx:Multiscale formulation}
The reduced radial migration model employed in the present work is based on the assumption that particle migration occurs over a timescale much longer than the oscillation period of the carrier flow (i.e., scale separated). In this appendix, we derive the cycle-averaged description in the small-Stokes-number limit.

We rewrite Eq.(\ref{eq:radial_MR}) in terms of the control parameter $\Pi=\Lambda/St$ in order to bring out the relative strengths of the competing transport mechanisms,
\begin{equation}
\frac{dR}{d\tau}=St\left[\underset{\circled{1}}{\underbrace{\frac{U_{\theta}^{2}\left(R\right)}{R}\cos^{2}\left(\Omega\tau\right)}}-\underset{\circled{2}}{\underbrace{\Pi\frac{d\ln\Theta}{dR}}}\right],
\label{eq_appdx:dr_dt}
\end{equation}
where $\Pi=\Lambda/St$ indicates the relative importance of thermophoretic and inertia-induced transport.
 
The present analysis focuses on forcing frequencies comparable to the characteristic vortex turnover rate, $\Omega=\omega/q \sim O(1)$, consequently the asymptotic ordering $\varepsilon\ll1$ is equivalent to $St\ll1$, as a result the expansion is expressed directly in powers of $St$. From Eq.(\ref{eq:Rexpansion}), we have,
\begin{equation}
R(\tau,\tau_s)
=
R_0(\tau_s)
+
St R_1(\tau,\tau_s)
+
O(St^2),
\label{eq_appdx:Rexpansion}
\end{equation}
where $\tau_s$ is the slow timescale and $\tau=qt$ is the fast timescale.  
\begin{equation}
    \mbox{Consider term }  \circled{1} = \frac{U_{\theta}^{2}\left(R\right)}{R}\cos^{2}\left(\Omega\tau\right), \\ \mbox{with } M(R)=\frac{U_{\theta}^2}{R}.
\end{equation}

Taylor expansion of $M(R)$ about $R_{0}$ gives,
\begin{equation}
M\left(R\right)=M\left(R_{0}\right)+\left(R-R_{0}\right)\left(\frac{dM}{dR}\right)_{R_{0}}+O\left(R-R_{0}\right)^{2}.
\end{equation}
Using Eq.(\ref{eq_appdx:Rexpansion}) and the definition of $M(R)$, we have,

\begin{equation}
M\left(R\right)=\frac{U_{\theta}^{2}\left(R_{0}\right)}{R_{0}}+StR_{1}\left(\tau_{s},\tau\right)\left(\frac{dM}{dR}\right)_{R_{0}}+O\left(R-R_{0}\right)^{2}.
\label{eq_appdx:M_R}
\end{equation}

\begin{equation}
\mbox{Similarly for term } \circled{2}=\Pi\frac{d\ln\Theta}{dR}, 
\mbox{with }    N\left(R\right)=\frac{d\ln\Theta}{dR}.    
\end{equation}

Taylor expansion of $N(R)$ about $R_{0}$ gives,

\begin{equation}
N\left(R\right)=N\left(R_{0}\right)+\left(R-R_{0}\right)\left(\frac{dN}{dR}\right)_{R_{0}}+O\left(R-R_{0}\right)^{2}.
\end{equation}

Using Eq.(\ref{eq_appdx:Rexpansion}) and the definition of $N(R)$, we have,

\begin{equation}
N\left(R\right)=\left(\frac{d\ln\Theta}{dR}\right)_{R_{0}}+StR_{1}\left(\tau_{s},\tau\right)\left(\frac{dN}{dR}\right)_{R_{0}}+O\left(R-R_{0}\right)^{2}.
\label{eq_appdx:N_R}
\end{equation}.

Substituting Eqs.(\ref{eq_appdx:M_R},\ref{eq_appdx:N_R}) into Eq.(\ref{eq_appdx:dr_dt}) and regrouping,

\begin{align}
    \frac{dR}{d\tau}=St\left[\frac{U_{\theta}^{2}\left(R_{0}\right)}{R_{0}}\cos^{2}\left(\Omega\tau\right)-\Pi\left(\frac{d\ln\Theta}{dR}\right)_{R_{0}}\right]+ \\ St^{2}\left[R_{1}\left(\tau_{s},\tau\right)\left(\frac{dM}{dR}\right)_{R_{0}}\cos^{2}\left(\Omega\tau\right)-\Pi R_{1}\left(\tau_{s},\tau\right)\left(\frac{dN}{dR}\right)_{R_{0}}\right].
    \label{eq_appdx:dR_dtau}
\end{align}

Application of the chain rule yields, 

\begin{equation}
\frac{dR\left(\tau_{s},\tau\right)}{d\tau}=\frac{\partial R}{\partial\tau}+St\frac{\partial R}{\partial\tau_{s}}.
\label{eq:chain_rule}
\end{equation}

Using Eq.(\ref{eq_appdx:Rexpansion}) in Eq.(\ref{eq:chain_rule}),  we have
\begin{equation}
\frac{dR\left(\tau_{s},\tau\right)}{d\tau}=St\left[\frac{\partial}{\partial\tau}\left(R_{1}\left(\tau_{s},\tau\right)\right)+\frac{dR_{0}}{d\tau_{s}}\right]+St^{2}\frac{\partial}{\partial\tau_{s}}R_{1}\left(\tau_{s},\tau\right).
\end{equation}

Substituting the above in the LHS of Eq.(\ref{eq_appdx:dR_dtau}) and grouping the $O(St)$ terms, Eq.(\ref{eq_appdx:dR_dtau}) becomes,

\begin{equation}
\frac{dR_{0}}{d\tau_{s}}+\frac{\partial}{\partial\tau}\left(R_{1}\left(\tau_{s},\tau\right)\right)=\frac{U_{\theta}^{2}\left(R_{0}\right)}{R_{0}}\cos^{2}\left(\Omega\tau\right)-\Pi\left(\frac{d\ln\Theta}{dR}\right)_{R_0}.
\end{equation}
On account of $St\ll1$, we ignore the $O(St^2)$ terms. Now defining the cycle average over one forcing period as,

\begin{equation}
\langle f\rangle=\frac{\Omega}{2\pi}\int_{0}^{2\pi/\Omega}f(\tau)d\tau
\end{equation}

Noting that $R_{1}(\tau_s,\tau)$ is periodic (see Eq.(\ref{eq:periodic})), its cycle average vanishes. It is easy to verify that $\langle \cos^{2}(\Omega\tau)\rangle = 1/2$, and since the thermophoretic term is time independent of time, we get,

\begin{equation}
\frac{dR_{0}}{d\tau_{s}}=\frac{U_{\theta}^{2}\left(R_{0}\right)}{2R_{0}}-\Pi\left(\frac{d\ln\Theta}{dR}\right)_{R_{0}}
\label{eq_appdx:slow_timescale_eq}
\end{equation}

 Eq.(\ref{eq_appdx:slow_timescale_eq}) is the radial drift equation in the slow timescale which is Eq.(\ref{eq:slowdrift}).

\section{Appendix: Robustness to the prescribed vortex and temperature profile}
\label{appdx:robustness}
The results presented in the main text were obtained using the stretched-exponential vortex profile
In this appendix, we demonstrate the robustness of the saddle node bifurcation as the mechanism of the loss of particle trapping to different velocity amplitude and temperature profiles. We begin by considering,   
\begin{equation}
U_\theta(R)=\exp(-R^a),
\end{equation}

with $a=4$, which leads to the cycle-averaged inertial drift

\begin{equation}
G(R)=\frac{1}{2}R\exp(-2R^a).
\end{equation}

To assess whether the trapping-loss transition depends on the particular choice of vortex model, the analysis was repeated using two additional azimuthal velocity distributions. The first is a Lamb--Oseen vortex,

\begin{equation}
U_\theta(R)=\frac{1-e^{-R^2}}{R},
\end{equation}

which produces the drift function

\begin{equation}
G(R)=\frac{1}{2}\frac{(1-e^{-R^2})^2}{R^3}.
\end{equation}

The second is a smoothed Rankine vortex,

\begin{equation}
U_\theta(R)=\frac{R}{1+R^2},
\end{equation}

for which

\begin{equation}
G(R)=\frac{1}{2}\frac{R}{(1+R^2)^2}.
\end{equation}

For each vortex model, the fixed points were obtained from the reduced radial dynamics

\begin{equation}
G(R)-\Pi\frac{d\ln\Theta}{dR}=0.
\end{equation}

\begin{figure}
\includegraphics[width=0.7\linewidth]{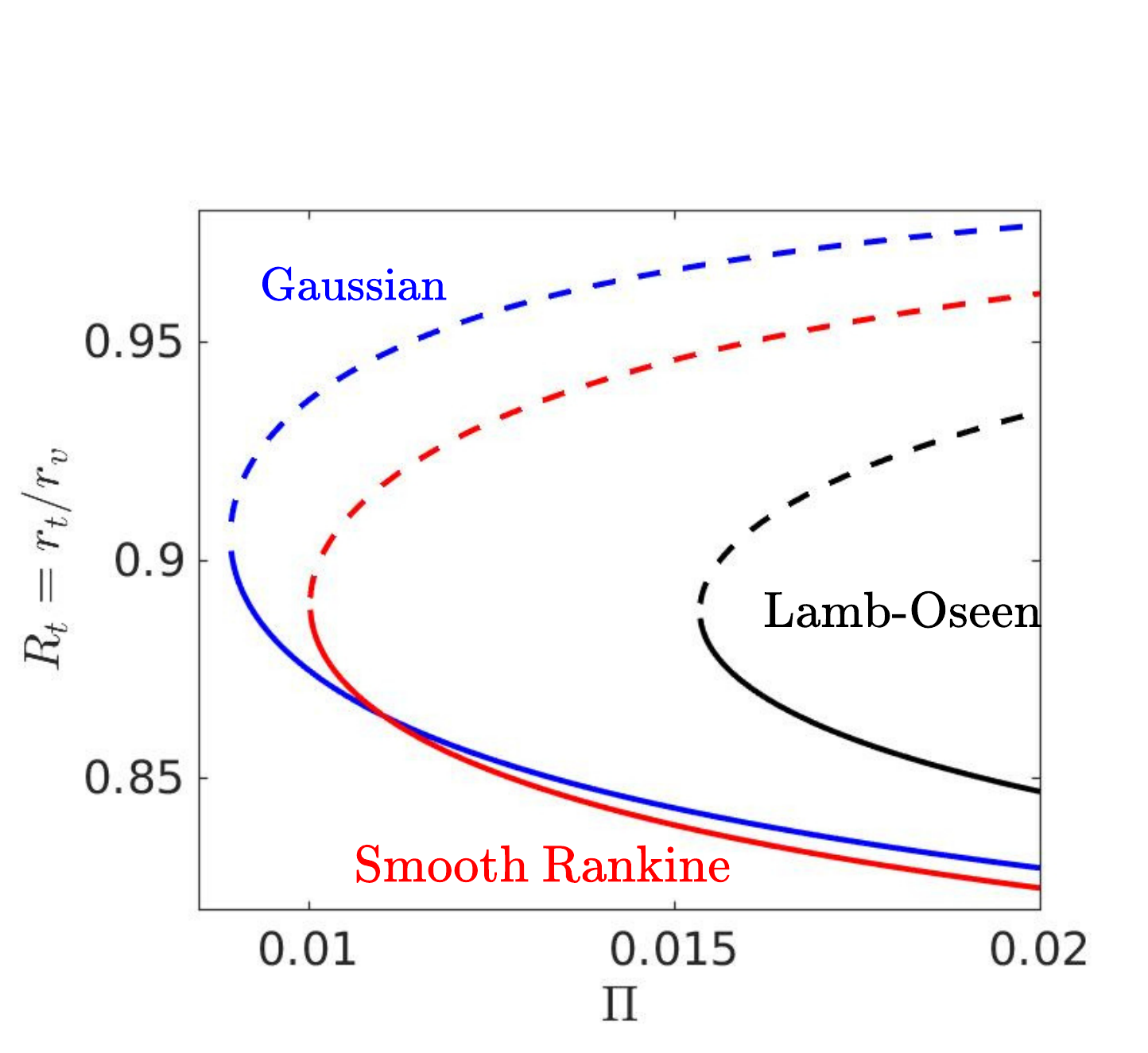}
\includegraphics[width=0.7\linewidth]{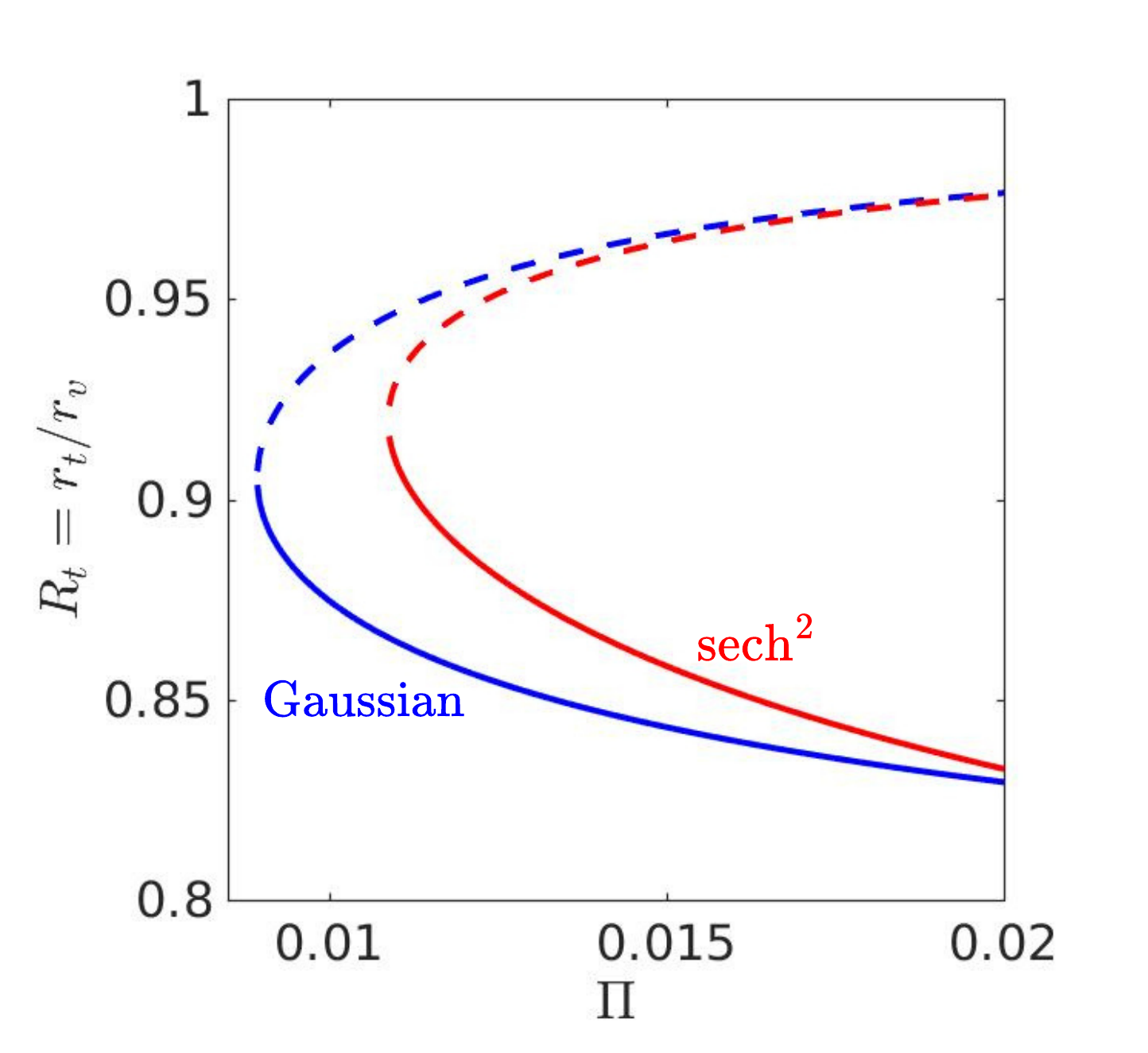}
\caption{(a) Robustness of the trapping-loss bifurcation to the vortex profile. Solid lines denote stable branches and dashed lines denote unstable branches. The saddle-node topology persists for stretched-exponential, Lamb--Oseen, and smoothed Rankine vortices. (b) Robustness of the trapping-loss bifurcation to the temperature profile. Solid lines denote stable branches and dashed lines denote unstable branches. The saddle-node topology persists for both Gaussian and Sech$^2$ annular thermal layers.}
\label{fig:bifurcation_robustness}
\end{figure}

The corresponding bifurcation diagrams are shown in Fig.~\ref{fig:bifurcation_robustness}. Despite the substantial differences in the radial structure of the three velocity fields, all cases exhibit the same qualitative behavior. For sufficiently large values of $\Pi$, two equilibrium solutions are present: a stable trapping state (solid branch) and an unstable fixed point (dashed branch). As $\Pi$ is reduced, the two branches approach one another and eventually merge at a finite critical value $\Pi_c$, beyond which no equilibrium trapping states exist.

The principal effect of modifying the vortex profile is therefore a quantitative shift in the critical parameter and trapping radius. The overall topology of the bifurcation diagram remains unchanged. In particular, the coexistence of stable and unstable branches and their finite-parameter coalescence persist for all three vortex models. This indicates that the trapping-loss transition is not a consequence of the specific stretched-exponential profile adopted in the main text, but rather reflects the generic competition between outward inertia-induced transport and inward thermophoretic migration.

The thermal field employed in the main text was represented by the Gaussian distribution

\begin{equation}
\Theta(R)
=
1+A_T
\exp\!\left[
-\frac{(R-R_s)^2}{\Delta^2}
\right].
\end{equation}

To examine the sensitivity of the results to the thermal-field model, the analysis was repeated using a localized hyperbolic-secant profile,

\begin{equation}
\Theta(R)
=
1+A_T
\operatorname{sech}^2
\!\left(
\frac{R-R_s}{\Delta}
\right),
\end{equation}

which possesses the same characteristic location and width as the Gaussian profile. In both cases, the fixed points were obtained from

\begin{equation}
G(R)-\Pi\frac{d\ln\Theta}{dR}=0,
\end{equation}

using the stretched-exponential vortex profile adopted in section \ref{sec:problem_formulation}.

The resulting bifurcation diagrams are shown in  Figs.(\ref{fig:bifurcation_robustness}a,b). The two temperature profiles produce quantitatively different trapping radii and critical parameter values. Nevertheless, the qualitative structure of the dynamics remains unchanged. For both thermal fields, the reduced dynamics admits a stable trapping state and an unstable separatrix that approach one another and ultimately coalesce at a finite critical value $\Pi_c$. The persistence of this branch-merging behavior demonstrates that the trapping-loss transition is insensitive to the detailed functional form of the temperature field.

\nocite{*}

\bibliography{apssamp}

\end{document}